\documentclass[sigconf, 9pt]{acmart}
\AtBeginDocument{%
  }

\acmConference[MOBISYS '26]{}{June 21 - 25, 2026}{Cambridge, UK}
\acmISBN{978-1-4503-XXXX-X/2018/06}

\begin{document}

\title{Scaling Mobile Agent Systems: From Capability Density to Collective Intelligence}


\author{Bowei He}
\affiliation{%
  \institution{Mohamed bin Zayed University of Artificial Intelligence}
  \country{}}
\email{Bowei.He@mbzuai.ac.ae}

\renewcommand{\shortauthors}{Bowei et al.}

\begin{abstract}
Mobile agent systems are emerging as a key paradigm for enabling intelligent applications on edge devices and in AIoT ecosystems. However, their scalability is fundamentally constrained by limited on-device computation and fragmented intelligence across devices. In this work, we propose a unified research agenda for scaling mobile agent systems along two complementary dimensions: (1) improving capability density of individual agents through compact foundation model design and compression, and (2) enabling collective intelligence via communication-rich multi-agent collaboration. Building on recent model and infrastructure advances, this vision aims to transform isolated mobile agents into a distributed intelligent system that is efficient and scalable.
\end{abstract}


\acmYear{2026}\copyrightyear{2026}
\setcopyright{cc}
\setcctype[4.0]{by}
\acmConference[MobiSys '26]{The 24th Annual International Conference on Mobile Systems, Applications and Services}{June 21--25, 2026}{Cambridge, United Kingdom}
\acmBooktitle{The 24th Annual International Conference on Mobile Systems, Applications and Services (MobiSys '26), June 21--25, 2026, Cambridge, United Kingdom}
\acmDOI{10.1145/3812835.3814812}
\acmISBN{979-8-4007-2711-5/26/06}



\maketitle

\section{Introduction}
Recent advances in agent foundation models~\cite{li2025chain} and harness systems~\cite{openclaw2025} have enabled increasingly capable intelligent assistants that can perform multi-step task execution. At the same time, there is a growing demand to bring such capabilities onto mobile and edge devices, driven by requirements in latency, privacy, and always-on availability in AIoT scenarios. Prior work on mobile agents~\cite{wen2025autodroid} and on-device LLM services~\cite{yin2025elastic} has demonstrated the feasibility of deploying intelligent agents directly on personal devices.

Despite this progress, scaling mobile agent systems remains fundamentally challenging. On one hand, mobile devices are constrained by limited computation, memory, and energy budgets, making it impractical to deploy high-capacity models, which often exceed 100B parameters. On the other hand, intelligence remains fragmented across devices, as agents typically operate in isolation and cannot effectively collaborate to solve complex tasks.

\begin{figure}[t]
    \centering
    \includegraphics[width=0.98\linewidth]{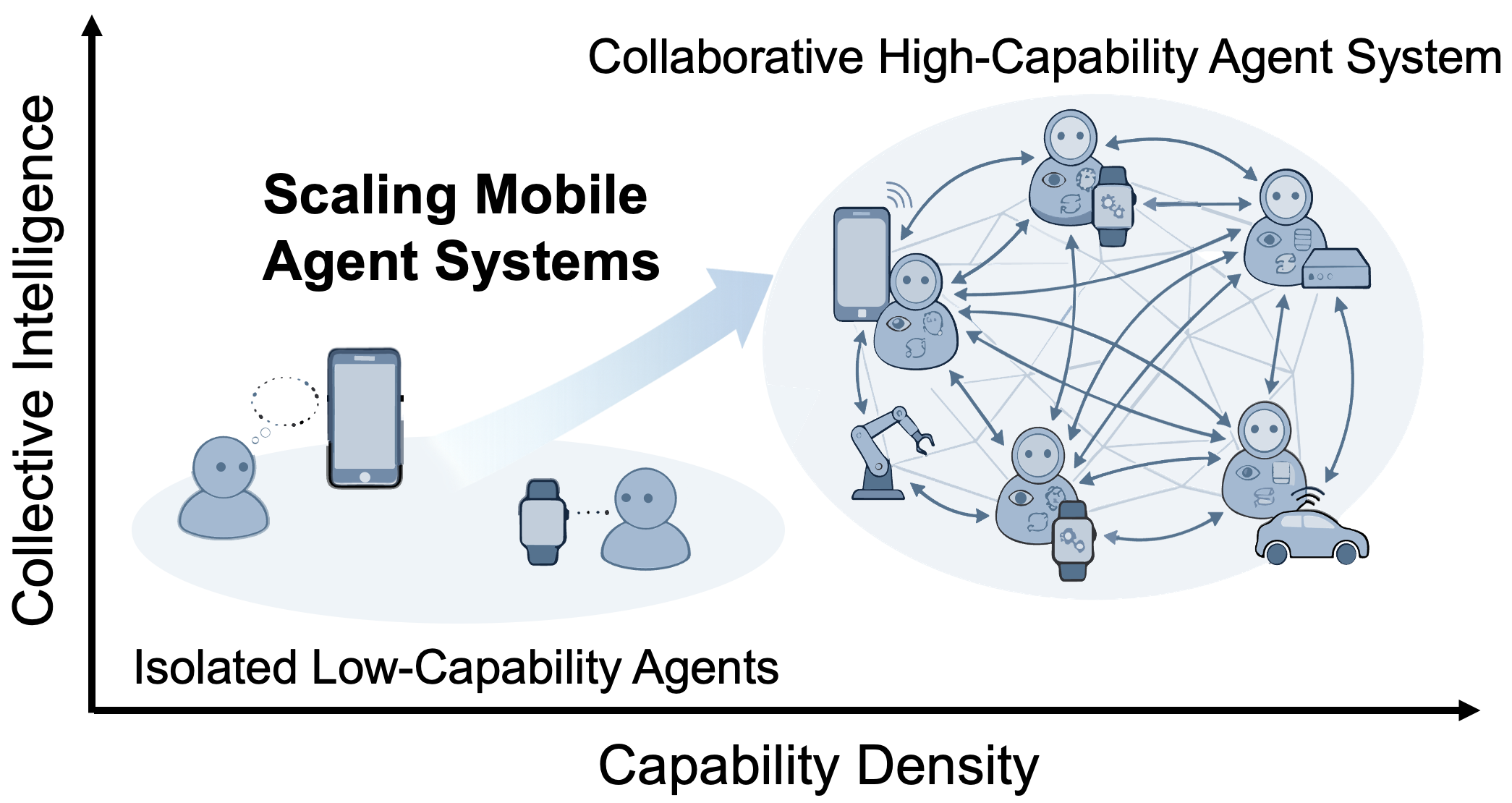}
    \caption{The visualization for scaling mobile agent systems along two complementary dimensions: improving capability density of individual agents and enabling collective intelligence via multi-agent collaboration.
    }
    \vspace{-3mm}
    \label{fig:scaling_mobile_agents}
\end{figure}

To address these limitations, we propose a \textit{two-dimensional} scaling framework for mobile agent systems, illustrated in Figure~\ref{fig:scaling_mobile_agents}. The first dimension focuses on improving the \textbf{capability density} of individual agents, maximizing the intelligence per unit resource on a single device. The second dimension focuses on enabling \textbf{collective intelligence} across multiple agents through communication, coordination, and routing. These two dimensions are complementary: advances in individual efficiency enable broader deployment, while advances in collaboration unlock system-level intelligence beyond any single agent. Together, these two directions provide a principled path toward scalable mobile intelligence.

\section{Scaling Capability Density}
Improving capability density aims to make each mobile agent as capable and efficient as possible within strict resource constraints. This requires rethinking both agent foundation model design and the surrounding system infrastructure.

A central challenge lies in adapting foundation models to resource-limited environments. Our work explores a combination of model compression techniques, including pruning~\cite{hepaser, liu2025optishear}, quantization~\cite{hepreserving}, distillation~\cite{hepedagogically}, and model merging~\cite{liu2025sens, liu20251bit}, to reduce model size and inference cost while preserving performance. These techniques have been studied both independently and in combination, demonstrating that careful and principled integration can significantly improve overall system efficiency without sacrificing capability.

Beyond architectural optimization, data also plays a critical role in maintaining model performance after compression. Our work shows that calibration data curation can substantially improve quantized models~\cite{hepreserving}, while pedagogically-inspired data synthesis enhances knowledge distillation~\cite{hepedagogically}. These results highlight the importance of data-centric approaches in enabling efficient models.

Another important aspect is robustness. Mobile agents must operate reliably under dynamic inputs and potentially adversarial environments. Our work on robustness certification provides guarantees for compressed models against backdoor and adversarial perturbations~\cite{hecertifying}, which is essential for real-world deployment.

Taken together, these efforts move toward compact yet powerful agent foundation models that retain strong reasoning and interaction abilities while operating efficiently on-device. Such agents serve as the core components for scalable mobile intelligence.
\vspace{-2mm}

\section{Scaling Collective Intelligence}
While improving individual agents is necessary, it is still not sufficient. Many real-world tasks exceed the capability of a single agent, especially under resource constraints on mobile or edge devices. This motivates the second dimension of scaling: enabling collective intelligence through collaboration among agents.

A key infrastructure of collective intelligence is communication. Traditional message-passing mechanisms are often inefficient or insufficient for complex coordination in large-scale multi-agent systems. Recent work on semantic communication protocols enables agents to exchange high-level information, such as intent, context, and intermediate reasoning states, significantly improving both richness and usefulness~\cite{protocol}. This represents a shift from low-level data exchange to meaning-aware interaction.

Coordination further amplifies agent capabilities. By decomposing tasks and assigning subtasks to specialized agents, systems can achieve performance beyond what any individual agent can provide. Our prior work demonstrates the effectiveness of structured coordination patterns, such as hierarchical actor–refiner framework for reasoning~\cite{he2026search} and function-oriented agent pipelines for domain-specific applications like drug design~\cite{gaocidd}.

Efficient routing is another critical challenge in multi-agent systems. Given a network of heterogeneous agents, the system must dynamically determine which agent is best suited for a given query. Our work on routing explores context-aware strategies that optimize both efficiency and performance, including analytical routing models~\cite{chen20261}, lightweight high-speed routing systems without dedicated GPUs~\cite{liu202698}, and adaptive routing for multimodal agents~\cite{liu2026adaptive}.

These components: semantic communication, structured coordination, and efficient routing, collectively enable a network of agents to serve as a coherent distributed system. In this setting, intelligence emerges not from a single model, but from the continuous interaction and coordination among many agents.
\vspace{-2mm}

\section{Conclusions and Future Works}
We present a unified perspective on scaling mobile agent systems by jointly advancing capability density at the device level and collective intelligence at the system level. A key future direction is to co-design model and infrastructure to improve agent efficiency. Another important path is to move beyond task-level collaboration toward persistent mobile agent ecosystems, where agents continuously evolve and coordinate under dynamic resource and network constraints. In such ecosystems, agents may maintain long-term memory, learn skills from repeated interactions, specialize over time, and form adaptive collaboration patterns.

\section*{Short Bio}
\textbf{Bowei He} is currently a Postdoctoral Researcher at MBZUAI, supervised by Prof. Xue (Steve) Liu. His recent research focuses on developing efficient on-device agent foundation models for mobile applications, as well as designing communication protocols and collaboration frameworks for agents. Besides, he is broadly interested in agent reinforcement learning, agent memory, long-horizon and lifelong evolving agents, and physical AI agents.

\bibliographystyle{ACM-Reference-Format}
\bibliography{reference}

\appendix

\end{document}